\documentclass[10pt]{article}
\usepackage[margin=1in]{geometry}
\usepackage{booktabs}
\usepackage{graphicx}
\usepackage{hyperref}
\usepackage{microtype}
\usepackage{amsmath}
\usepackage{xcolor}
\title{ArtifactBench: Lineage-Aware Evaluation of AI-Generated Music Detectors under Distribution Shift}
\author{Heewon Oh\\Intrect\\
\texttt{heewon.oh@intrect.io}\\
\href{https://orcid.org/0009-0000-2287-1753}{ORCID: 0009-0000-2287-1753}}
\date{}

\begin{document}
\maketitle

\begin{abstract}
AI-generated music detectors are commonly compared using aggregate scores on
benchmarks whose training overlap, generator lineage, source provenance, and
audio-transformation history are only partially observable. This paper introduces
ArtifactBench, a lineage-aware evaluation suite for measuring detector behavior
across generator families and versions, real-music domains, collection-cohort
shift, and inference coverage. The benchmark groups source recordings and their derived
variants by content identity, separates calibration from final testing, records
inference failures independently from classification errors, and reports
source-level performance with uncertainty in addition to aggregate metrics. We
evaluate multiple publicly available detectors under a version-pinned common
protocol and examine how leakage control, cohort availability, threshold policy,
and model-specific missingness alter measured performance and model ranking.
On the 562-track common-success test intersection, ArtifactNet obtains 0.982
AUROC and 0.918 balanced accuracy, compared with 0.761/0.776 for the public
Deezer detector; SpecTTTra and CLAM fall below 0.30 AUROC under this shifted
cohort.  These results also expose substantial generator- and real-domain shifts
that aggregate scores alone conceal.
\end{abstract}

\section{Introduction}
AI-generated music detection is increasingly evaluated under distribution shift:
new generators, changing model versions, post-processing, and real music drawn
from domains unlike the detector's training data.  Recent datasets have made the
task measurable at useful scale~\cite{rahman2025sonics,batra2025mom,pascu2026echoes},
while detector studies have repeatedly shown that high in-domain accuracy can
coexist with weak robustness to unseen generators or audio
transformations~\cite{afchar2025challenges,afchar2025fourier}.  A less visible
source of variation is the evaluation cohort itself.  A benchmark name may refer
to several manifests, only a subset of referenced audio may still be available,
and a model's public training split may overlap the nominal test set.  If each
adapter silently skips a different set of files, the resulting aggregate scores
do not compare the same population.

This paper studies AI-music detection as an evaluation-governance problem.  We
introduce ArtifactBench v2, a lineage-aware protocol that binds every public
track identifier to a content digest, source family, generator version when
known, and a lineage identifier.  It separates cohort construction from model
execution, freezes calibration, validation, and test partitions before final
analysis, and records decoding and inference failures separately from wrong
predictions.  A common operating point is selected on calibration data alone;
threshold-free and model-native results remain visible so that conclusions do
not depend on one threshold convention.

Our contributions are:
\begin{itemize}
  \item a forensic audit of three incompatible ArtifactBench manifest lineages
  and the audio bytes that can actually be reconstructed;
  \item a content- and lineage-controlled 828-track primary cohort spanning 13
  synthetic source strata and two real-music domains, split into calibration,
  validation, and sealed test partitions without lineage conflicts;
  \item a version-pinned runner that applies one deterministic cohort to four
  public detectors, preserves raw track-level scores, and treats inference
  coverage as a reported result; and
  \item an analysis of source and generator-version shift, threshold policy, upstream split exposure,
  and their effect on measured performance and model ranking.
\end{itemize}

ArtifactBench v2 is not proposed as the largest training corpus, nor as a claim
that every track is unknown to every representation model.  Its goal is narrower:
to make the population, provenance, operating point, and failure denominator of
each comparison auditable.

\section{Related Work}
\subsection{AI-Generated Music Detection}
SONICS introduced a large full-song corpus and the SpecTTTra family of
spectrogram transformers~\cite{rahman2025sonics}.  Deezer demonstrated that a
detector trained from artificial reconstructions can obtain very high held-out
accuracy while emphasizing manipulation and unseen-generator
risks~\cite{afchar2025challenges}; later work connected detection evidence to
Fourier structure~\cite{afchar2025fourier}.  MoM and CLAM combine music and
speech-oriented representations under contrastive training~\cite{batra2025mom}.
ArtifactNet instead operates on estimated codec residuals and harmonic/percussive
structure~\cite{oh2026artifactnet}.  MusicDET frames the task as real-only
one-class density estimation~\cite{han2026musicdet}.  These methods differ not
only in architecture but also in crop duration, aggregation, training corpora,
and score calibration; comparing published headline values therefore conflates
model and protocol.

\subsection{Audio Deepfake Benchmarks}
Echoes controls semantic alignment between bona-fide references and generated
tracks to reduce shortcut learning~\cite{pascu2026echoes}.  HAIM expands the
label space beyond binary origin to stages of human--AI production
interaction~\cite{go2026haim}.  BAMM shows a further domain gap in real television
broadcast audio~\cite{lopezayala2026bamm}.  These datasets expose complementary
axes---semantic alignment, production provenance, and deployment channel.  Our
focus is the measurement layer shared across such axes: track identity,
train--test exposure, missingness, threshold selection, and paired comparison.

\subsection{Dataset Lineage, Leakage, and Distribution Shift}
Content leakage is stronger than a duplicated filename and weaker than an exact
byte match alone.  Re-encoding, trimming, or alternate containers may preserve
the same recording while changing its digest.  ArtifactBench therefore uses
both a cryptographic content identity and a lineage group.  The split unit is
the lineage, not the file.  We also distinguish (i) exact content overlap,
(ii) membership in a published upstream train or validation partition, and
(iii) unresolved exposure where the public training procedure does not provide
an identity map.  These states are reported rather than collapsed into a binary
``unseen'' label.

\section{ArtifactBench Design}
\subsection{Scope and Intended Use}
The primary task is binary track-level detection of fully generated music versus
human-produced music.  The benchmark is intended for comparative research,
failure analysis, and reproducibility testing.  It does not establish authorship,
copyright status, intent, or whether a partially AI-assisted production should
be labeled synthetic.  Segment localization, hybrid-production tracking, and
commercial adjudication are outside the primary task.

\subsection{Source Provenance and Licensing}
The primary cohort contains 605 synthetic and 223 real tracks.  The synthetic
side includes 310 tracks from the AIME release~\cite{grotschla2025aime}, 180
Suno CDN or supplementary tracks, and 115 Udio CDN or supplementary tracks.
The real side contains 150 tracks recovered from the official FMA large archive
and 73 hard negatives referenced by stable source identifiers.  FMA metadata
was matched for all 150 selected IDs~\cite{defferrard2017fma}.

Because source licenses differ and redistribution rights for platform-collected
audio are not uniformly established, the v2 public package is metadata-first:
it contains path-free identifiers, digests, source provenance, split assignments,
and reconstruction instructions, but does not introduce a new audio bundle.
The private runtime manifest binds those public identities to locally verified
bytes and is not released.

\subsection{Content Identity and Lineage Groups}
Each row contains a stable \texttt{track\_id}, binary label, source stratum,
generator family and version when available, SHA-256 of the evaluated bytes, and
a \texttt{lineage\_id}.  Exact byte matches against recovered ArtifactNet
training audio were removed.  Two real-audio variants found by acoustic
fingerprinting were also removed.  A lineage may contain several containers or
transformations, but no lineage may cross protocol partitions.

\subsection{Generator, Domain, Time, and Codec Axes}
Table~\ref{tab:cohort} summarizes the primary population.  AIME sources cover
open and hosted generation systems available during its collection period;
the supplementary Suno and Udio strata add later or independently collected
outputs.  FMA provides artist-licensed 30-second real music, whereas the second
real stratum emphasizes web-sourced hard negatives.  These labels are retained
in every raw prediction to prevent aggregate performance from hiding a source
collapse.

\begin{table}[t]
\centering
\caption{Frozen primary cohort before protocol partitioning.}
\label{tab:cohort}
\begin{tabular}{lrr}
\toprule
Source group & Strata & Tracks \\
\midrule
AIME generators & 9 & 310 \\
Suno supplementary/CDN & 2 & 180 \\
Udio supplementary/CDN & 2 & 115 \\
FMA real hard negatives & 1 & 150 \\
Web real hard negatives & 1 & 73 \\
\midrule
Total & 15 & 828 \\
\bottomrule
\end{tabular}
\end{table}

\subsection{Calibration, Validation, and Sealed Test Partitions}
We assign lineages deterministically within source strata using a published salt
and target proportions of 20/10/70.  This yields 166 calibration tracks (121 AI,
45 real), 83 validation tracks (61 AI, 22 real), and 579 sealed test tracks
(423 AI, 156 real).  The split was written to the manifest before the final
four-model analysis.  Calibration selects operating thresholds; validation is
diagnostic only and cannot trigger retuning; test is used once for the reported
comparison.

\section{Evaluation Protocol}
\subsection{Version-Pinned Model Adapters}
We evaluate ArtifactNet v9.4, SpecTTTra-$\alpha$-120s, the public Deezer ISMIR
2025 logistic detector, and CLAM.  Repository, model, and feature-encoder
revisions are pinned rather than fetched from mutable default branches.  ONNX
models are materialized with adjacent external-data files and their revisions
are recorded in each run manifest.  CLAM additionally records the checkpoint
SHA-256 and the pinned MERT and Wav2Vec2 encoder revisions.

\subsection{Common and Model-Native Preprocessing}
All adapters receive the same decoded mono waveform and selected track IDs in
the same deterministic order.  Resampling occurs inside each adapter to its
native sampling rate.  Fixed-duration baselines use a center crop, not a random
crop.  SpecTTTra uses its 120-second input policy; CLAM uses a deterministic
90-second crop.  ArtifactNet evaluates seven evenly spaced four-second chunks
and uses the median of finite chunk scores when at least four of seven chunks
are valid.  We retain a strict sensitivity result in which any non-finite chunk
causes a track failure and publish the chunk-level failure log.  Four is the
smallest strict majority; this source-agnostic reliability rule was fixed before
the sealed-split analysis and was not optimized against labels or performance.
No model-specific replacement track is introduced after a decode or inference
failure.

\subsection{Threshold Policies}
The primary operating threshold for each model is chosen using calibration rows
only.  Among observed score thresholds with false-positive rate (FPR) no greater
than 5\%, we maximize true-positive rate (TPR); ties select the lower threshold.
The threshold is then frozen for validation and test.  We separately report the
native 0.5 threshold and threshold-free area under the ROC curve (AUROC) and
average precision (AUPRC).  This separation prevents a favorable calibration
choice from being presented as representation quality.

\subsection{Inference Failures and Missing Data}
Resolution, decode, model inference, non-finite output, and out-of-range
probability failures are distinct events.  They are not converted into a class
prediction and are not silently dropped from the attempted denominator.  Main
tables show both successfully scored tracks and failures.  Pairwise comparisons
use the common successfully scored test-ID intersection; coverage itself remains
a model outcome.

\subsection{Metrics and Statistical Uncertainty}
For each partition and source we report TPR or FPR, and for mixed-label
partitions AUROC, AUPRC, F1, balanced accuracy, and the confusion matrix.
Ninety-five percent intervals are computed with 2,000 label-stratified lineage
bootstrap replicates.  Resampling the lineage rather than the file prevents
multiple encodings of one recording from being treated as independent evidence.

\section{Experiments}
\subsection{Models and Reproducibility Environment}
All final inference runs are executed on one Apple Mac Studio (M1 Max, 32 GB)
from a frozen runtime copy whose 828 content hashes were reverified.  PyTorch
models use MPS where supported; the ArtifactNet ONNX model uses CPU execution.
Each result directory contains the input-manifest digest, selected track IDs,
seed, package versions, platform string, model provenance, per-track scores,
and structured failures.

\subsection{Leakage and Cohort Audit}
The public v1.0.1 manifest contains 6,200 rows and a 2,280-row test partition;
the public v1.1 branch contains a 2,224-row purged test set; and a local
post-purge manifest contains 6,014 actual rows despite metadata describing
6,183.  An earlier eight-model comparison scored only 2,104 tracks because 120
referenced FMA files were HTML responses rather than audio.  We recovered all
150 selected FMA tracks from the official archive and verified their identities.

The lineage-clean intermediate cohort contains 2,236 rows.  Against public
SONICS lineage, 241 rows map to train or validation, 454 to test only, and 1,541
are external.  Against public MoM lineage, 200 map to its training partition,
197 to its test partition, 1,523 are external, and 316 real rows remain
unresolved because the released CLAM training code applies a seeded random split
after local embedding-availability filtering without publishing the realized
identity map.  The conservative 828-track primary cohort excludes all native
SONICS- and MoM-derived source families.  We still avoid describing it as
universally training-unseen.

\section{Results}
All values in this section are generated from the frozen per-track JSON bundle.
No number is transcribed from the earlier ArtifactNet paper or the 2,104-track
exploratory comparison.

\subsection{Paired Aggregate Performance}
Table~\ref{tab:aggregate} reports classification metrics on the 562 test tracks
successfully scored by all four models.  The scored and coverage columns retain
the full 579-track attempted denominator.  ArtifactNet leads the paired
comparison with AUROC 0.982, AUPRC 0.994, and balanced accuracy 0.918.  The
public Deezer detector is second at 0.761 AUROC and 0.776 balanced accuracy.
SpecTTTra and CLAM obtain AUROC 0.299 and 0.284, respectively, showing that their
score orderings do not transfer to this cohort.  For CLAM, satisfying the
calibration FPR constraint requires the next representable threshold above its
maximum score of 1.0; the resulting operating point predicts no positives.  We
therefore show its threshold-free and native-threshold behavior rather than
interpreting the calibrated F1 in isolation.

\begin{table*}[t]
\centering
\caption{Paired test performance. Metric columns use the 562-track common-success
intersection; coverage uses all 579 attempted test tracks.  Thresholds were
selected on calibration data only.}
\label{tab:aggregate}
\resizebox{\textwidth}{!}{\begin{tabular}{lrrrrrrrrr}
\toprule
Model & scored & $\tau$ & AUROC & AUPRC & F1 & BAcc & TPR & FPR & coverage \\
\midrule
ArtifactNet v9.4 & 562/579 & 0.9909 & 0.982 & 0.994 & 0.923 & 0.918 & 0.865 & 0.029 & 0.971 \\
SpecTTTra-$\alpha$ & 579/579 & 0.9695 & 0.299 & 0.688 & 0.159 & 0.526 & 0.087 & 0.036 & 1.000 \\
Deezer ISMIR & 579/579 & 0.3165 & 0.761 & 0.925 & 0.758 & 0.776 & 0.624 & 0.072 & 1.000 \\
CLAM & 579/579 & 1.0000 & 0.284 & 0.647 & 0.000 & 0.500 & 0.000 & 0.000 & 1.000 \\
\bottomrule
\end{tabular}
}
\end{table*}

The lineage-bootstrap intervals in Table~\ref{tab:intervals} preserve the same
ordering.  ArtifactNet's AUROC interval is 0.971--0.991, whereas Deezer's is
0.723--0.798.  The non-overlap is descriptive evidence for this frozen cohort,
not a universal claim over future generators.

\begin{table}[t]
\centering
\caption{Ninety-five percent lineage-bootstrap intervals (2,000 replicates) on
the paired test intersection.}
\label{tab:intervals}
\resizebox{\linewidth}{!}{\begin{tabular}{lrrrr}
\toprule
Model & AUROC & AUPRC & TPR & FPR \\
\midrule
ArtifactNet v9.4 & 0.971--0.991 & 0.989--0.997 & 0.832--0.898 & 0.007--0.058 \\
SpecTTTra-$\alpha$ & 0.245--0.352 & 0.662--0.714 & 0.061--0.116 & 0.007--0.072 \\
Deezer ISMIR & 0.723--0.798 & 0.912--0.938 & 0.579--0.671 & 0.029--0.115 \\
CLAM & 0.236--0.333 & 0.623--0.675 & 0.000--0.000 & 0.000--0.000 \\
\bottomrule
\end{tabular}
}
\end{table}

\subsection{Generator-Version and Real-Domain Shift}
Table~\ref{tab:groups} separates source cohorts without treating collection
membership as a precise per-track timestamp.  ArtifactNet retains TPR 0.926 on
AIME and 0.897 on supplementary Suno, but falls to 0.650 on supplementary Udio.
Deezer shows the opposite pattern: TPR is 0.332 on AIME but rises to 0.968 on
Suno and 0.875 on Udio.  Its real-music FPR changes from 0.020 on FMA to 0.211
on the web hard-negative cohort.  Thus neither the generator side nor the real
side is represented by a single stable domain.

\begin{table}[t]
\centering
\caption{Cohort-group rates on the common-success test intersection at each
model's calibration-selected threshold.}
\label{tab:groups}
\resizebox{\linewidth}{!}{\begin{tabular}{lrrrrr}
\toprule
Model & AIME TPR & Suno TPR & Udio TPR & FMA FPR & Web FPR \\
\midrule
ArtifactNet v9.4 & 0.926 & 0.897 & 0.650 & 0.040 & 0.000 \\
SpecTTTra-$\alpha$ & 0.097 & 0.103 & 0.037 & 0.050 & 0.000 \\
Deezer ISMIR & 0.332 & 0.968 & 0.875 & 0.020 & 0.211 \\
CLAM & 0.000 & 0.000 & 0.000 & 0.000 & 0.000 \\
\bottomrule
\end{tabular}
}
\end{table}

\begin{figure*}[t]
\centering
\includegraphics[width=\textwidth]{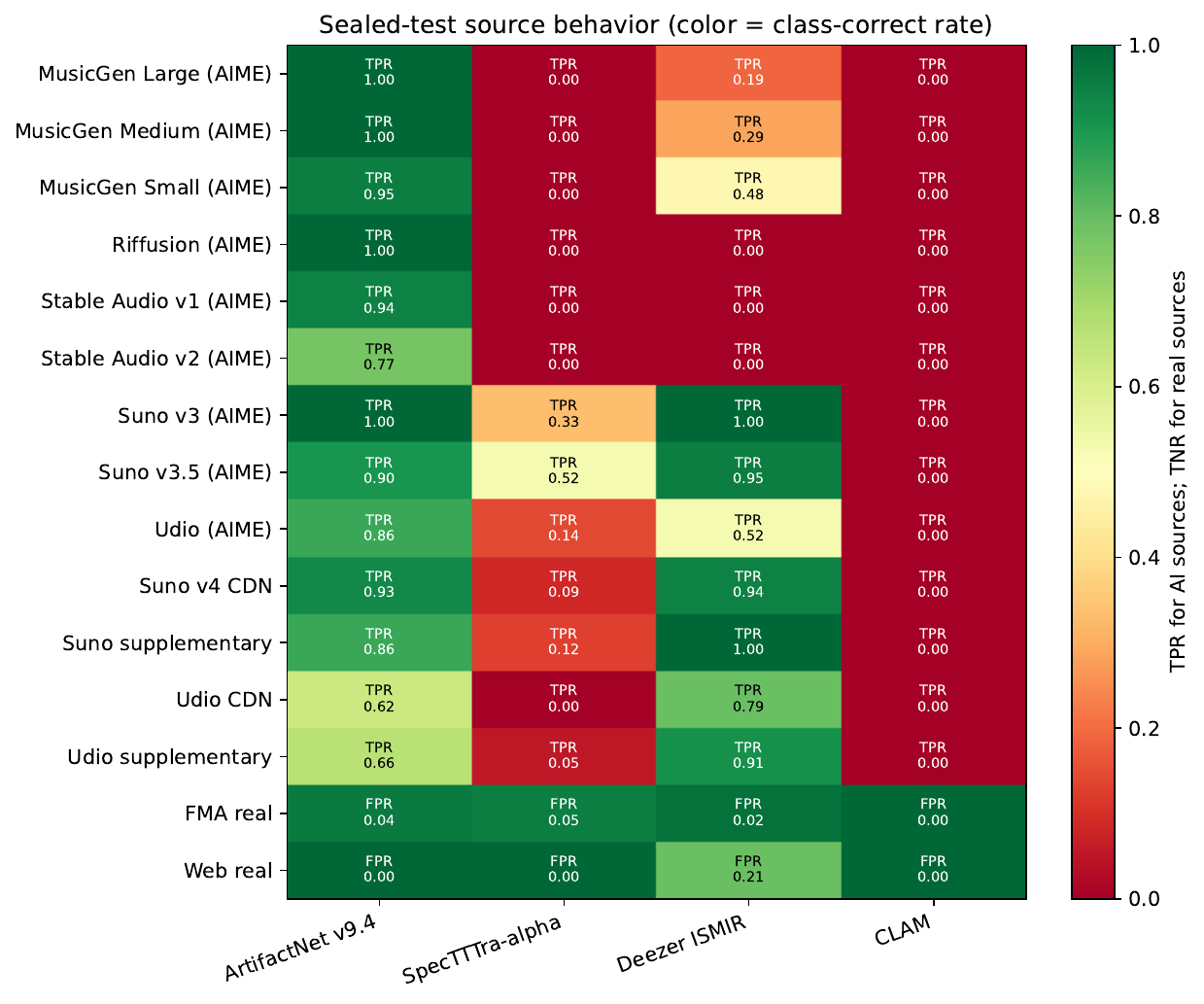}
\caption{Per-source TPR for synthetic strata and FPR for real strata on the
paired test intersection.  Values use calibration-selected thresholds.}
\label{fig:source}
\end{figure*}

Figure~\ref{fig:source} gives the corresponding source-stratum view.  It makes
clear that a favorable group mean can coexist with a generator-version collapse;
the labels denote collection cohorts and known generator versions, not exact
generation dates.

\subsection{Coverage and Numerical Reliability}
The three comparison baselines score all 579 test tracks.  ArtifactNet scores
562 (97.1\% coverage) under the declared finite-chunk rule, leaving 17 test
failures.  Across all 828 tracks, 165 contain at least one non-finite chunk; 141
retain at least four of seven finite chunks and are recovered by the predeclared
majority rule, leaving 24 unresolved tracks.  Non-finite chunks are not confined
to silence: their measured RMS ranges up to $-4.69$ dBFS.  This points to a
numerical model-path failure rather than a simple silent-input condition.

Table~\ref{tab:coverage} applies an intentionally pessimistic classification
bound in which every missing synthetic track is a false negative and every
missing real track is a false positive.  ArtifactNet's F1 then falls from 0.923
on common successful tracks to 0.904, and its balanced accuracy to 0.865.  This
bound does not fabricate probabilities, so AUROC and AUPRC remain undefined for
failed rows.

\begin{table}[t]
\centering
\caption{Coverage and failure-as-error classification bound on all 579 test
attempts.}
\label{tab:coverage}
\resizebox{\linewidth}{!}{\begin{tabular}{lrrrrrrr}
\toprule
Model & scored & failures & coverage & F1 & BAcc & TPR & FPR \\
\midrule
ArtifactNet v9.4 & 562/579 & 17 & 0.971 & 0.904 & 0.865 & 0.865 & 0.135 \\
SpecTTTra-$\alpha$ & 579/579 & 0 & 1.000 & 0.159 & 0.528 & 0.087 & 0.032 \\
Deezer ISMIR & 579/579 & 0 & 1.000 & 0.756 & 0.777 & 0.624 & 0.071 \\
CLAM & 579/579 & 0 & 1.000 & 0.000 & 0.500 & 0.000 & 0.000 \\
\bottomrule
\end{tabular}
}
\end{table}

Under the stricter sensitivity policy that rejects a track after any non-finite
chunk, ArtifactNet scores only 467 test tracks.  On those tracks it obtains
AUROC 0.985 and balanced accuracy 0.932, but failure-as-error balanced accuracy
drops to 0.710 because 112 test attempts are unresolved.  The finite-chunk rule
therefore improves coverage materially without creating a better score for any
failed chunk; both policies and the chunk-level log are released.

\subsection{Ranking Sensitivity}
ArtifactNet and Deezer rank first and second under calibrated balanced accuracy,
native-threshold balanced accuracy, and AUROC (Table~\ref{tab:threshold}).  The
lower two positions change: CLAM ranks above SpecTTTra at the native 0.5
threshold, while SpecTTTra ranks above CLAM after calibration and by AUROC.
More importantly, native 0.5 behavior is operationally poor for both: SpecTTTra
has balanced accuracy 0.355, and CLAM 0.473 with TPR 0.889 but FPR 0.942.  A
single headline F1 would conceal that CLAM's native score is driven by predicting
most tracks as synthetic.

\begin{table}[t]
\centering
\caption{Ranking sensitivity on the paired test intersection.}
\label{tab:threshold}
\resizebox{\linewidth}{!}{\begin{tabular}{lrrr}
\toprule
Model & calibrated BAcc & native-0.5 BAcc & AUROC \\
\midrule
ArtifactNet v9.4 & 0.918 & 0.934 & 0.982 \\
SpecTTTra-$\alpha$ & 0.526 & 0.355 & 0.299 \\
Deezer ISMIR & 0.776 & 0.767 & 0.761 \\
CLAM & 0.500 & 0.473 & 0.284 \\
\bottomrule
\end{tabular}
}
\end{table}

\subsection{Paired Codec Robustness}
Codec-pair experiments are secondary because the frozen primary run contains no
derived codec variants.  A later codec analysis must use identical excerpts and
retain the parent lineage; it is not mixed into the primary 579-track test.

\clearpage
\section{Discussion}
\subsection{What Aggregate F1 Hides}
An aggregate F1 score combines class prevalence, one operating threshold, and a
particular mixture of generator and real domains.  A detector can therefore
appear strong while failing one real-music stratum, or weak when its ranking
scores are useful but its default threshold is miscalibrated.  Source-level
rates and threshold-free metrics are necessary to distinguish these cases.

\subsection{Benchmark Construction Bias}
Benchmark curation can inadvertently reward a detector through codec, duration,
sampling-rate, or source shortcuts.  Semantic alignment addresses one part of
this problem~\cite{pascu2026echoes}; content lineage and split exposure address
another.  Neither guarantees deployment validity.  The broadcast degradation
reported by BAMM~\cite{lopezayala2026bamm} illustrates why a clean-file benchmark
must state its domain rather than imply universal detection.

\subsection{Limitations and Non-Claims}
The primary cohort is modest, class-imbalanced, and concentrated in several
generator families.  Platform version labels are incomplete for supplementary
tracks.  Public lineage evidence cannot reconstruct every CLAM real-training
identity, and content hashes cannot detect every perceptual derivative.  The
benchmark evaluates fully generated versus real music, not the degree or stage
of AI assistance.  Finally, the public research implementation of the Deezer
method is not Deezer's production detector~\cite{afchar2025fourier}.

\section{Data and Code Availability}
The version-pinned runner, public manifest, raw model scores, structured failure
logs, analysis scripts, and checksums are released through
\href{https://github.com/Intrect-io/artifactbench}{GitHub} and
\href{https://huggingface.co/datasets/intrect/artifactbench}{Hugging Face}.  The release is
metadata-first and does not add a new audio bundle; each source must be acquired
under its upstream license and verified against the published digest.

\section{Ethics, Licensing, and Data Governance}
Detector errors can harm creators if scores are treated as proof of authorship or
fraud.  ArtifactBench is intended to measure systems, not adjudicate individual
works.  Per-track public results use benchmark IDs rather than local paths or
unnecessary creator-identifying strings.  The release does not add a new bundle
of platform- or YouTube-sourced audio; users retrieve source data under its own
terms.  FMA licenses remain attached per track, and non-commercial restrictions
on SONICS, Deezer research code, and other upstream assets are preserved.

\section{Conclusion}
ArtifactBench v2 makes the denominator of AI-music detector evaluation explicit.
By binding scores to content and lineage identity, freezing threshold selection
before sealed testing, and reporting coverage separately from classification,
it turns several hidden benchmark choices into inspectable artifacts.  The
result is not a universal test of synthetic music, but a reproducible way to
locate where detectors generalize, where they fail, and which conclusions change
when the cohort or operating point changes.

\section*{Relationship to Prior Work}
ArtifactNet~\cite{oh2026artifactnet} introduced a residual-based detector and used
an earlier ArtifactBench snapshot as one evaluation set. The present work treats
the detector only as one baseline and contributes a lineage-aware benchmark design,
audited release artifacts, and a comparative analysis of evaluation sensitivity.

\bibliographystyle{plain}
\bibliography{refs}
\end{document}